\documentclass[manuscript]{aastex631}
\usepackage{amsmath}
\usepackage{romannum}
\usepackage{graphics}
\usepackage{natbib}
\usepackage{amsmath}
\usepackage{amssymb}
\usepackage{longtable,booktabs}
\usepackage{array, makecell}

\begin{document}
\title{Turbulence, Waves, and Taylor's Hypothesis for Heliosheath Observations}
\author{L.-L. Zhao}
\affiliation{Center for Space Plasma and Aeronomic Research (CSPAR), The University of Alabama in Huntsville, Huntsville, AL 35805, USA}
\affiliation{Department of Space Science, The University of Alabama in Huntsville, Huntsville, AL 35805, USA}
\author{G. P. Zank}
\affiliation{Center for Space Plasma and Aeronomic Research (CSPAR), The University of Alabama in Huntsville, Huntsville, AL 35805, USA}
\affiliation{Department of Space Science, The University of Alabama in Huntsville, Huntsville, AL 35805, USA}
\author{M. Opher}
\affiliation{Center for Space Physics, Boston University, Boston, MA, USA}
\author{B. Zieger}
\affiliation{Center for Space Physics, Boston University, Boston, MA, USA}
\author{H. Li}
\affiliation{Los Alamos National Laboratory, Los Alamos, NM 87545, USA}
\author{V. Florinski}
\affiliation{Center for Space Plasma and Aeronomic Research (CSPAR), The University of Alabama in Huntsville, Huntsville, AL 35805, USA}
\affiliation{Department of Space Science, The University of Alabama in Huntsville, Huntsville, AL 35805, USA}
\author{L. Adhikari}
\affiliation{Center for Space Plasma and Aeronomic Research (CSPAR), The University of Alabama in Huntsville, Huntsville, AL 35805, USA}
\affiliation{Department of Space Science, The University of Alabama in Huntsville, Huntsville, AL 35805, USA}
\author{X. Zhu}
\affiliation{Center for Space Plasma and Aeronomic Research (CSPAR), The University of Alabama in Huntsville, Huntsville, AL 35805, USA}
\author{M. Nakanotani}
\affiliation{Department of Space Science, The University of Alabama in Huntsville, Huntsville, AL 35805, USA}
\begin{abstract}
Magnetic field fluctuations measured in the heliosheath by the Voyager spacecraft are often characterized as compressible, as indicated by a strong fluctuating component parallel to the mean magnetic field. However, the interpretation of the turbulence data faces the caveat that the standard Taylor's hypothesis is invalid because the solar wind flow velocity in the heliosheath becomes subsonic and slower than the fast magnetosonic speed, given the contributions from hot pickup ions in the heliosheath. We attempt to overcome this caveat by introducing a 4D frequency-wavenumber spectral modeling of turbulence, which is essentially a decomposition of different wave modes following their respective dispersion relations. Isotropic Alfv\'en and fast mode turbulence are considered to represent the heliosheath fluctuations. We also include two dispersive fast wave modes derived from a three-fluid theory. We find that (1) magnetic fluctuations in the inner heliosheath are less compressible than previously thought. An isotropic turbulence spectral model with about 25\% in compressible fluctuation power is consistent with the observed magnetic compressibility in the heliosheath; (2) the hot pickup ion component and the relatively cold solar wind ions induce two dispersive fast magnetosonic wave branches in the perpendicular propagation limit. Pickup ion fast wave may account for the spectral bump near the proton gyrofrequency in the observable spectrum; (3) it is possible that the turbulence wavenumber spectrum is not Kolmogorov-like although the observed frequency spectrum has a -5/3 power-law index, depending on the partitioning of power among the various wave modes, and this partitioning may change with wavenumber.
\end{abstract}

\section{Introduction}\label{sec:introduction}

The solar wind plasma interacts with the interstellar medium and creates the heliospheric bubble \citep{Parker1961}. Within the heliosphere, the supersonic and super-Alfv\'enic solar wind expands until it reaches the heliospheric termination shock (HTS) where the solar wind flow decelerates and becomes subsonic. The inner heliosheath is the region between the HTS and the heliopause (HP) \citep{Zank1999SSRv, Zank2015}.
The Voyager 1 and 2 spacecraft entered the inner heliosheath in 2004 and 2007, respectively.
Both Voyagers measured turbulent magnetic field fluctuations during their journey across the heliosheath \citep[e.g.,][]{Burlaga2008,Burlaga2012, Fraternale2019}.
The turbulence in the inner heliosheath is often characterized as compressible, as suggested by the observed comparable parallel and perpendicular fluctuations with respect to the mean magnetic field \citep[e.g.,][]{Burlaga2006,Richardson2013,Burlaga2014, Fraternale2019, Zhao2019ApJb}.

An important yet underappreciated caveat of turbulence measurements in the heliosheath is that the standard Taylor's hypothesis is expected to be invalid.
In the simplest terms, Taylor's hypothesis converts the observed timescale into a length scale based on the flow velocity relative to the observer, i.e., frequency $\Omega$ into wavenumber $k$, i.e., $k = \Omega / U$ \citep{Taylor1938}.
The underlying assumption is that the flow velocity $U$ is much larger than the characteristic propagation speed of the fluctuations.
According to measurements made by the Voyager PLS and LECP instruments, the flow velocity in the heliosheath is about 150$\,$km/s or less \citep[e.g.,][]{Richardson2014, Cummings2021},  which is slower than the typical fast magnetosonic speed (e.g., $>\,200\,$km/s) due to the dominant pickup ion (PUI) pressure in the heliosheath \citep[e.g.,][]{Zank2018PUI}. This is of course expected for the downstream fluid in the shock frame. The Voyager spacecraft speed is $\sim\,$17$\,$km/s and is ignored in this paper.
Despite this fact, Taylor's hypothesis is still used in many investigation of Voyager observations in the heliosheath.
An argument that may be made for using Taylor's hypothesis is that the flow velocity is approximately perpendicular to the mean magnetic field.
Since Alfv\'en and slow magnetosonic waves reduce to non-propagating structures in the perpendicular propagation limit, it may be argued that the flow velocity is still larger than the wave speed.
However, this is not a sufficient justification for Taylor's hypothesis, the reason being that waves do not propagate perpendicular to the mean magnetic field exclusively. The implicit application of Taylor's hypothesis can lead to an incorrect interpretation of the Voyager turbulence data.
In the inner heliosphere, e.g., near Earth orbit, the solar wind is typically a low-$\beta$ (ratio between thermal and magnetic pressure) plasma. The ``2D + slab'' model suggests that turbulence is dominated by the nonpropagating 2D structures, whose wavevector is perpendicular to the mean magnetic field. This is based on the nearly incompressible magnetohydrodynamic (MHD) theory \citep[e.g.,][]{Zank1992JGR,Zank1993PF,Zank2017ApJ} in the $\beta \ll 1$ or $\beta \sim 1$ limit. The heliosheath is a high-beta plasma ($\beta \gg 1$) due to the dominant PUI thermal pressure and thus turbulence in the heliosheath is not expected to have a dominant 2D component due to the weak background magnetic field.
For parallel propagating waves, although the wave propagation speed such as the Alfv\'en speed is $\sim\,$50$\,$km/s, slower than the typical flow velocity of 150$\,$km/s, it can still be comparable or larger than the flow velocity parallel to the mean magnetic field.
To summarize, Taylor's hypothesis is expected to break down for the Voyager observations in the heliosheath due to two reasons: (\romannum1) the flow velocity is slower than the speed of fast magnetosonic waves propagating in all directions; and (\romannum2) the parallel flow velocity may also be slower than the Alfv\'en wave speed.

To overcome the limitation of Taylor's hypothesis, \citet{Zhao2024} presented a method based on models of the frequency-wavenumber spectrum (or 4D $\omega$-$\boldsymbol{k}$ spectrum) of turbulence and applied the method to the inner heliosphere, especially for regions close to the Sun where the solar wind speed is comparable to the Alfv\'en speed. Their results show that the full 4D description of turbulent fluctuations including the effects of nonzero frequency has important consequences for the 
spacecraft observed 1D reduced turbulence spectrum.
In this work, a similar method is applied to the heliosheath observations. There are some important distinctions from the previous work. Specifically, (\romannum1) fast magnetosonic modes and Alfv\'en modes are both considered in the 4D spectral modeling while \cite{Zhao2024} considered Alfv\'en waves and nonpropagating structures; (\romannum2) the wavenumber spectrum is assumed to be isotropic due to the high-beta environment of the heliosheath; and (\romannum3) two dispersive fast wave mode branches from the multi-fluid model of non-equilibrated pickup ions, solar wind ions, and electrons are included in the 4D spectrum modeling. The paper is organized as follows. In Section \ref{sec:method}, we introduce the 4D turbulence spectral model applicable to the heliosheath. In Section 3, we discuss the comparison between the derived 1D reduced observable spectra and Voyager observations in the heliosheath. Section 4 provides a summary and discussion.

\section{Method}\label{sec:method}

As shown by \citet{Zhao2024} and \citet{Fredricks1976}, the observed turbulence power spectrum can be related to the 4D frequency-wavenumber spectrum via
\begin{equation}\label{eq:4Dspectra}
  P_{obs}(\Omega) = \int P(\omega, \boldsymbol{k}) \delta(\Omega - \omega - \boldsymbol{k}\cdot\boldsymbol{U}) d^3 k d\omega,
\end{equation}
where $\Omega$ is the observed frequency (in the instrument frame), $\omega$ is the fluctuation frequency in the plasma flow frame, $\boldsymbol{U}$ is the flow velocity relative to the spacecraft,
and $\boldsymbol{k}$ represents the wavevector measured in the plasma flow frame.
Intuitively, Equation \eqref{eq:4Dspectra} means that the observed fluctuation at a given frequency $\Omega$ is a superposition of wave modes with various wavevectors and frequencies being Doppler shifted to the observed frequency. 

Based on Equation \eqref{eq:4Dspectra}, the key to understand the observed spectrum $P_{obs}({\Omega})$ is to model the 4D spectrum $P(\omega, \boldsymbol{k})$. The standard Taylor's hypothesis essentially means $P(\omega, \boldsymbol{k}) = P(\boldsymbol{k})$ and neglects the fluctuation intrinsic frequency.
However, \citet{Zhao2024} showed that the full 4D spectrum including the effects of nonzero frequency has important consequences for the observation of the turbulence spectrum when the wave speed is comparable to the flow speed.
Here, we follow the method and model the 4D frequency-wavenumber spectrum by convolution of the 3D wavenumber spectrum $P(\boldsymbol{k})$ and the frequency response function $F(\omega, \boldsymbol{k})$,
\begin{equation}\label{eq:pwk}
P(\omega, \boldsymbol{k}) = P(\boldsymbol{k})F(\omega, \boldsymbol{k}).
\end{equation}
The frequency response $F(\omega, \boldsymbol{k})$ can incorporate the effects of both nonlinear broadening and the wave dispersion relations. The sweeping model of \citet{Kraichnan1964} suggests a Gaussian broadening function \citep{Bourouaine2018}, while recent numerical simulations suggest a Lorentzian broadening function \citep{yuen2023}. In these scenarios, the frequency response function $F(\omega, \boldsymbol{k})$ can be characterized by either a Gaussian or Lorentzian function, respectively, to describe the broadening of the 4D power spectrum around the wave resonance frequency (or zero for non-propagating modes). 
Previous studies suggest that frequency broadening is usually a small correction to the prediction of the observed spectrum, which affects only the spectrum near the spectral break, and the general spectral shape remains unchanged \citep[e.g.,][]{Narita2017, Bourouaine2019, Zhao2024}. Given the uncertainty associated with the 3D wavenumber spectral model $P(\boldsymbol{k})$ and dispersion relations of the dispersive waves in the heliosheath, it is safe to assume that the issue of frequency broadening is of secondary importance in the problem considered here. Thus, we neglect frequency broadening in this paper and take the frequency response function $F(\omega, \boldsymbol{k})$ to be a Dirac delta function $\delta(\omega - \omega_0(\boldsymbol{k}))$ with the wave frequency $\omega_0$ determined by its dispersion relation $\omega_0(\boldsymbol{k})$.   
The 4D power spectrum $P(\omega, \boldsymbol{k})$ is thereby a decomposition of different wave modes following their respective dispersion relations.

Voyager observations in the inner heliosheath suggest a strong magnetic compressibility, represented by the power ratio between the parallel fluctuations and perpendicular fluctuations \citep[e.g.,][]{Fraternale2019}. We consider the MHD Alfv\'en and fast magnetosonic waves to quantitatively investigate the compressibility observed by Voyager 1 and 2. 
The MHD slow modes are neglected in this study as they are heavily damped and may not exist as propagating waves in high-beta collisionless plasma \citep[e.g.,][]{Zank2014, majeski2023}. However, compressible mirror modes may replace slow modes in the perpendicular propagation limit when the solar wind ion temperature anisotropy $T_\perp/T_\parallel > 1$. It is indeed possible that mirror modes do contribute to the compressive fluctuations observed in the heliosheath. However, for simplicity, they are not included in the present spectral modeling and will be considered in the future. 
To model the 4D turbulence spectrum $P(\omega, \boldsymbol{k})$, one has to determine the 3D spatial spectral model. The common ``2D+slab'' model of turbulence in the inner heliosphere is not appropriate for the high-beta heliosheath since the magnetic field is not strong enough to warrant the use of a dominant 2D component. In this work, we consider an isotropic turbulence model for simplicity due to the weak background magnetic field. We use the convention that the mean magnetic field $\boldsymbol{B}_0$ is in the $z$-direction, the bulk flow velocity $\boldsymbol{U}$ is in the $x$-$z$ plane, and the $y$-axis completes the right-hand triplet. Isotropic turbulence is assumed in the sense that $P(\boldsymbol{k})$ depends only on the magnitude of the wavevector, $|\boldsymbol{k}| = \sqrt{k_x^2 + k_y^2 + k_z^2}$ and $k_x$, $k_y$, and $k_z$ are the three components of the wavevector. 

Therefore, the 4D frequency-wavenumber power spectrum for the isotropic turbulence can be described as,
\begin{equation}\label{eq:4A}
	P(\omega, \boldsymbol{k}) = P(\boldsymbol{k}) \delta(\omega - \omega_0(\boldsymbol{k})),\\
\end{equation}
where 
\begin{eqnarray}\label{eq:3Dspatial}
&& P(\boldsymbol{k}) =
\left\{
\begin{aligned}
& P_0 (|\boldsymbol{k}|/k_0)^{-\alpha}
, & \qquad  &
|\boldsymbol{k}| > k_0 \\ 
& P_0, &    & \textrm{otherwise}.
\end{aligned}
\right. 
\end{eqnarray}
Here, $P_0$ determines the spectral power for normalization and $\alpha$ represents the spectral index in wavenumber space. The bendover wavenumber $k_0$ corresponds to the correlation scale of turbulence spectrum. 
The wavenumber power spectrum $P(\boldsymbol{k})$ contains the sum of the diagonal components $P_{xx}$, $P_{yy}$, and $P_{zz}$ of the power spectral density (PSD) tensor, which represent the power contained in the fluctuations along the three coordinate axes $x$, $y$ and $z$.
For isotropic Alfv\'enic turbulence, $\omega_0(\boldsymbol{k}) = |k_z V_A|$, where $V_A$ is the Alfv\'en speed, and the 3D spectral density contains the incompressible components only, i.e., $P(\boldsymbol{k}) \sim P_{xx}+P_{yy}$.
The flow-frame frequency $\omega$ is set to be always positive while the wavevector $\boldsymbol{k}=(k_x, k_y, k_z)$ can be in any direction, meaning that counter-propagating Alfv\'en modes with equal strength are included as well.
Although non-propagating structures are not explicitly considered in our spectral model, they are still present in the isotropic Alfv\'enic turbulence because the dispersion relation $k_z V_A$ suggests that Alfv\'en waves reduce to zero-frequency non-propgating modes in the perpendicular wavevector limit. The magnetic fluctuations are decoupled from the velocity fluctuations in such limit and may be interpreted as ``magnetic islands'' \citep{Zank2021ApJ, Zank2023, zank2024}. Thus, the magnetic power of the isotropic Alfv\'enic turbulence considered here is non-zero when propagating perpendicularly, and is considered as a zero-frequency mode in our calculations.
We can then derive the predicted 1D frequency spectrum of the isotropic Alfv\'enic turbulence component using Equations \eqref{eq:4Dspectra} and \eqref{eq:4A}:
\begin{eqnarray}\label{eq:4F}
P^A_{obs}(\Omega) & = & \int P(\boldsymbol{k})\delta(k_x U_x + k_z U_z + |k_z V_A| - \Omega) d^3 k \nonumber\\
		& = & \int \frac{1}{|U_x|} P\left(|\boldsymbol{k}| = \sqrt{(k_z U_z + |k_z V_A| - \Omega)^2/U_x^2 + k_y^2 + k_z^2} \right) dk_y dk_z.
\end{eqnarray}

The second component we consider in the heliosheath is isotropic fast-mode turbulence, which can be modeled similarly with the MHD fast mode dispersion relation, i.e., $\omega_0(\boldsymbol{k}) = |\boldsymbol{k}| V_f$, where the fast mode wavevector $\boldsymbol{k}=(k_x, k_y, k_z)$. The square of the MHD fast speed $V^2_f = (V_A^2 + C_i^2 + ((V_A^2 + C_i^2)^2 - 4V_A^2 C_i^2 \cos^2\theta_{\boldsymbol{k}\boldsymbol{B}_0})^{1/2})/2$, where $C_i$ represents the sound speed of ions including both PUIs and solar wind ions. $\boldsymbol{B}_0$ is along the $z$-direction, so $\cos\theta_{\boldsymbol{k}\boldsymbol{B}_0} = k_z/\sqrt{k_x^2+k_y^2+k_z^2}$. 
The corresponding observable frequency spectrum for MHD fast mode is calculated from 
\begin{equation}\label{eq:6F}
	P^F_{obs}(\Omega) = \int P(\boldsymbol{k})\delta(k_x U_x + k_z U_z + |\boldsymbol{k}| V_f - \Omega) d^3 k.
\end{equation}
The integral here can then be reduced to a 2D integral as follows,
\begin{equation}
	P^F_{obs}(\Omega) = \int \frac{1}{|\partial(k V_f)/\partial k_y|} P\left(k_x, k_y(k_x, k_z, \Omega), k_z \right) dk_x dk_z.
\end{equation}
One has to be careful in choosing the integration limits of $k_x$ and $k_z$, because for a given observable frequency $\Omega$, there may be no solution for $k_y$ if $k_x$ and $k_z$ are too large. 
Technical details about the computation of the integral are shown in the Appendix.

Besides the standard MHD waves within a single-fluid plasma, \citet{Zieger2015,Zieger2020} also derived the dispersion relation of warm multi-fluid plasma for perpendicular wave propagation. The three-fluid model includes a low-temperature ion component that represents the relatively ``cold'' solar wind ions, a higher-temperature ion component that represents the hotter PUIs, and a electron component.
The thermal velocity of the relatively ``cold'' solar wind ions is much smaller than that of the PUIs. Note that the PUIs here represent particles in the energy range of $1-10\,$keV, while suprathermal particles with energies $>10\,$keV are neglected. In the perpendicular wavenumber limit, the MHD fast magnetosonic wave is then split into a high-frequency fast mode (HFF) that propagates in the hotter PUIs and a low-frequency fast mode (LFF) propagating in thermal solar wind ions.
While \cite{Zieger2015} considered a Maxwellian distributed (isotropic) PUI fluid, a more general description of PUI-mediated plasma waves, including collisionless heat conduction and viscosity due to pitch-angle scattering, was proposed by \cite{Zank2014}, where dispersion curves for the outer heliosphere ($>$10 AU), inner heliosheath, and very local interstellar medium (VLISM) were presented. The role of the nearly isotropic PUI distribution and the role of PUI heat flux in the damping of the wave modes were discussed. An 11th-order polynomial dispersion relation is obtained in \cite{Zank2014} and it is much too complicated to be included for the analysis presented here. However, we note that the dispersion curves of two separate PUI and solar wind ion fast modes, when propagating perpendicularly in the heliosheath, are similar in both \cite{Zank2014} and \cite{Zieger2015}. For simplicity, we use the dispersion relations of two fast magnetosonic mode waves derived in \cite{Zieger2015}.
An interesting feature of the multi-fluid fast modes is that they are dispersive waves, meaning that their propagation speed depends not only on the wavevector $\boldsymbol{k}$ direction, but also on its magnitude.
Dispersive waves can cause the spacecraft observed frequency spectrum to deviate from the power-law shape of the wavenumber spectrum, as suggested by \citet{Zieger2020}.
In their three-fluid model, the dispersion relation for perpendicular fast magnetosonic modes can be expressed as the solutions to the quadratic equation of $\omega_0^2$ \citep{Zieger2015}. 
\begin{equation}\label{eq:dispersion}
  A_2(k)\omega_0^4 - A_1(k)\omega_0^2 + A_0(k) = 0,
\end{equation}
where 
\begin{equation}
	A_2(k) = \sum_j \omega^2_{pj}\left(c^2k^2 + \sum_j \omega^2_{pj}\right),
\end{equation}

\begin{eqnarray}
	A_1(k) & = & \left(\sum_j \omega^2_{pj} \sum_j \omega_{gj}^{\ast 2}-\sum_j \omega^2_{pj} \omega_{gj}^{\ast 2} \right) \left(c^2k^2+\sum_j \omega_{pj}^2\right) \nonumber \\
	& - & \left[ \omega^2_{p\mathrm{P}} \omega^2_{p\mathrm{S}} (\omega_{g\mathrm{P}} -\omega_{g\mathrm{S}})^2 + \omega^2_{p\mathrm{S}} \omega^2_{p\mathrm{e}} (\omega_{g\mathrm{S}} -\omega_{g\mathrm{e}})^2+ \omega^2_{p\mathrm{e}} \omega^2_{p\mathrm{P}} (\omega_{g\mathrm{e}} -\omega_{g\mathrm{P}})^2\right],
\end{eqnarray}

\begin{eqnarray}
	A_0(k) & = & \left(\omega^2_{p\mathrm{P}} \omega_{g\mathrm{S}}^{\ast 2}\omega_{g\mathrm{e}}^{\ast 2} + \omega^2_{p\mathrm{S}} \omega_{g\mathrm{e}}^{\ast 2}\omega_{g\mathrm{P}}^{\ast 2} + \omega^2_{p\mathrm{e}} \omega_{g\mathrm{P}}^{\ast 2}\omega_{g\mathrm{S}}^{\ast 2} \right) \left(c^2k^2+\sum_j \omega_{pj}^2\right) \nonumber \\
	& - & \left[ \omega^2_{p\mathrm{P}} \omega^2_{p\mathrm{S}} \omega_{g\mathrm{e}}^{\ast 2}(\omega_{g\mathrm{P}} -\omega_{g\mathrm{S}})^2 + \omega^2_{p\mathrm{S}} \omega^2_{p\mathrm{e}}\omega_{g\mathrm{P}}^{\ast 2} (\omega_{g\mathrm{S}} -\omega_{g\mathrm{e}})^2+ \omega^2_{p\mathrm{e}} \omega^2_{p\mathrm{P}} \omega_{g\mathrm{S}}^{\ast 2}(\omega_{g\mathrm{e}} -\omega_{g\mathrm{P}})^2\right].
\end{eqnarray}
Here, $\omega_{g\mathrm{j}}^{\ast 2} = \omega_{gj}^2 + k c_j^2$. $\omega_{gj}$, $\omega_{pj}$, and $c_j$ are the gyrofrequency, the plasma frequency, and the sound speed of species $j$ (solar wind ion, pickup ion, and electron), and $c$ represents the speed of light. The two solutions of $\omega_0^2$ from Equation \eqref{eq:dispersion} represent two branches of fast wave modes, both of which are dispersive on fluid scales.
Note that Equation \eqref{eq:dispersion} is for perpendicular waves only ($k_\parallel = 0$), i.e., $k$ is the wave number perpendicular to the magnetic field in the dispersion relation \eqref{eq:dispersion}. 

Once we know the dispersion relations of HFF and LFF modes, one has to consider a suitable spatial spectral model to construct the 4D frequency-wavenumber spectra. 
Since the two fast mode waves from \cite{Zieger2015} model consider perpendicular wavenumber only, the 4D spectrum for HFF and LFF branches can be modeled as
$P(\omega, \boldsymbol{k}) = G(\boldsymbol{k}_\perp) \delta(\omega - \omega_{pui,\,sw})$. Here, $\omega_{pui}(|\boldsymbol{k}_\perp|)$ and $\omega_{sw}(|\boldsymbol{k}_\perp|)$ represent the HFF mode frequency and LFF frequency, respectively, which can be obtained from Equation \eqref{eq:dispersion}. We assume the wavenumber spectrum $G(\boldsymbol{k}_\perp)$ to be a broken power law for both HFF and LFF branches: 
\begin{eqnarray}\label{eq:multi3D} 
&& G(\boldsymbol{k}_\perp) =
\left\{
\begin{aligned}
& P_j (|\boldsymbol{k}_\perp|/k_0)^{-\alpha}
, & \qquad  &
|\boldsymbol{k}_\perp| > k_0 \\ 
& P_j, &    & \textrm{otherwise},
\end{aligned}
\right. 
\end{eqnarray}
where $P_j$ determines the spectral power of the fast mode driven by species $j$, PUI and solar wind ion;
$\alpha$ represents the spectral index in $\boldsymbol{k}$-space.
Therefore, the predicted frequency spectrum to be observed is calculated from 
\begin{eqnarray}\label{eq:18F} 
&& P^{F}_{pui}(\Omega) = \int G(\boldsymbol{k}_\perp)\delta(k_x U_x + \omega_{pui} - \Omega) dk_x dk_y; \\
&& P^F_{sw}(\Omega) = \int G(\boldsymbol{k}_\perp)\delta(k_x U_x + \omega_{sw} - \Omega) dk_x dk_y. \nonumber
\end{eqnarray}
The integral in Equations \eqref{eq:4F}, \eqref{eq:6F}, and \eqref{eq:18F} can be evaluated numerically and the technical details of the numerical integration are included in the Appendix.


\section{Results}
\subsection{Voyager 2 observations in the heliosheath}
\begin{figure}
\centering
\includegraphics[width=0.75\linewidth]{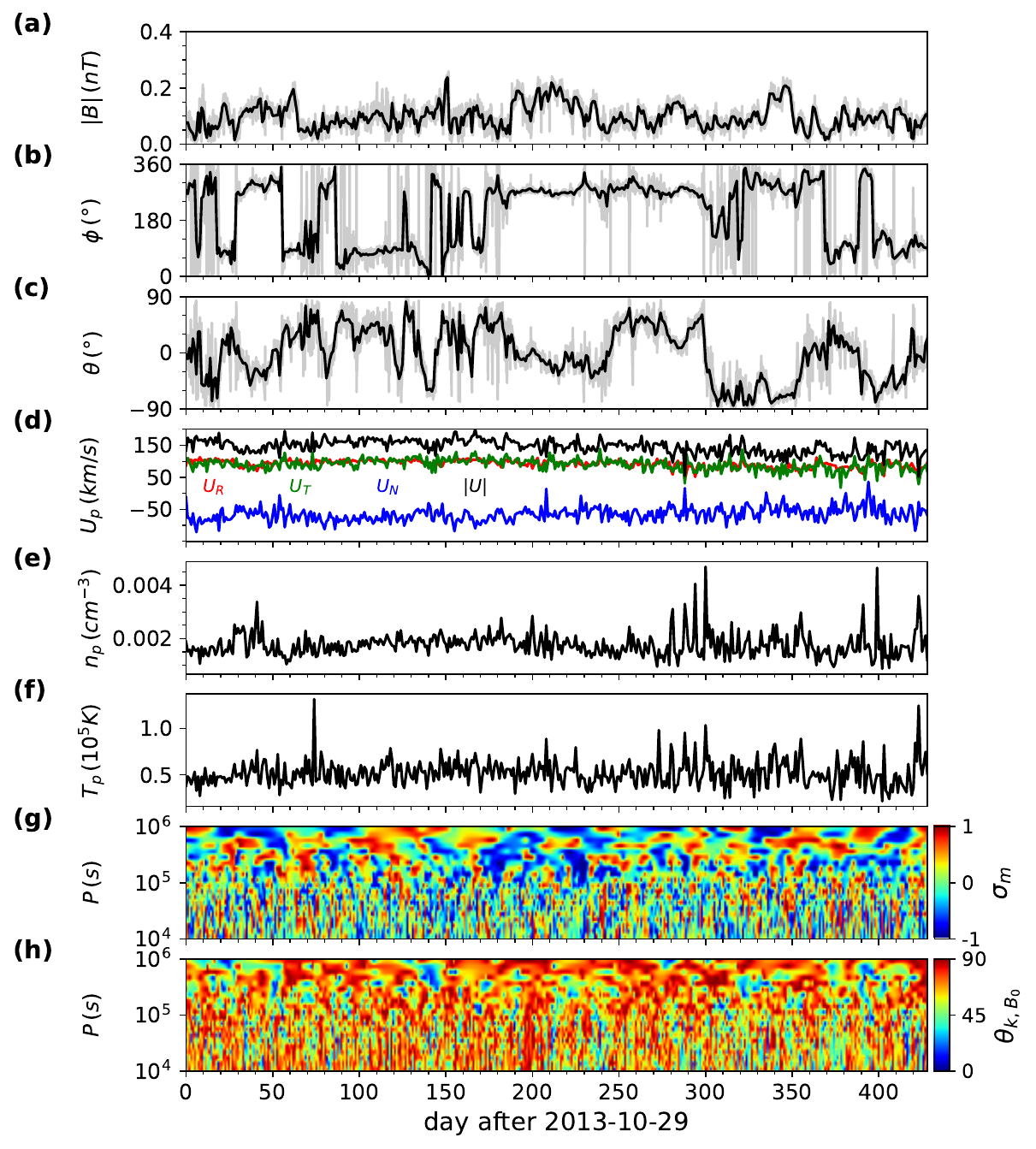}
	\caption{Overview of the Voyager 2 magnetic field and solar wind proton measurements in the inner heliosheath during the period from 2013-10-29 to 2015-01-01. Panels (a)-(c) show the daily averaged (black lines) magnetic field magnitude $|B|$, the azimuthal $\phi$ and the elevation $\theta$ angles of the magnetic field. 48 s resolution magnetic field data is shown as gray lines. Panel (d) shows the daily averaged solar wind proton velocity $\boldsymbol{U}_p$ in the $RTN$ reference frame, with the radial velocity $U_R$ in red, the tangential velocity $U_T$ in green, the normal speed $U_N$ in blue, and the total speed in black. Panels (e) and (f) show solar wind proton density $n_p$ and temperature $T_p$, respectively. Panels (g) and (h) show the wavelet spectrogram of the normalized reduced magnetic helicity $\sigma_m$ and the angle between wavevector $\boldsymbol{k}$ and the local mean magnetic field $\theta_{k, B_0}$, respectively. }\label{fig:time}
\end{figure}

To go beyond the standard Taylor's hypothesis, we apply the 4D frequency-wavenumber spectral modeling method to Voyager observations in the inner heliosheath. We select an interval in the inner heliosheath observed by Voyager 2 to illustrate how the observed fluctuations can be decomposed into a combination of wave modes with different power through spectral modeling. Figure \ref{fig:time} displays the daily averaged magnetic field magnitude $|B|$, azimuthal $\phi$ and elevation $\theta$ angles of the magnetic field directions, solar wind proton speed $\boldsymbol{U}_p$, density $n_p$, and temperature $T_p$ measured by Voyager 2 during the period from 2013 Oct 29 to 2015 Jan 1. The data are not uniform in temporal resolution and the highest resolution for magnetic field measurements is 48 seconds (grey lines in the top three panels). In panel (g), we show the wavelet spectrogram of the normalized reduced magnetic helicity $\sigma_m$ based on 48-second magnetic field measurements. $\sigma_m$ is calculated from the two perpendicular magnetic fluctuation components, i.e., $\sigma_m = {2 \operatorname{Im}(\delta \tilde{\boldsymbol{B}}_\mathrm{\perp 1}^* \delta \tilde{\boldsymbol{B}}_\mathrm{\perp 2})}/{\mathrm{Tr}(\boldsymbol{B})}$, where the tilde represents wavelet-transformed quantities. $\mathrm{Tr}(\boldsymbol{B})$ is the magnetic trace spectrum. From the figure, there are no clear signatures of a particular wave pattern during the interval, as the spectrogram does not have a certain period of time dominated by relatively large positive or negative $\sigma_m$ values associated with wave polarization with respect to the background magnetic field \citep{Zhao2021MHD, Zhao2021B}. Panel (h) shows the scale-dependent $\theta_{k, B_0}$, i.e., the angle between the wavevector $\boldsymbol{k}$ and the local mean magnetic field. The wavevector $\boldsymbol{k}$ is estimated by the singular value decomposition (SVD) method \citep{Santolik2003}. However, it should be noted that 48-second resolution magnetic field data are only available for very short periods of time, resulting in large data gaps. We simply linearly interpolate through the data gaps and discard the high-frequency part (i.e., period $p\le10^4\,$s) of the spectrogram in panels (g) and (h). The low-frequency range can be recovered well by linear interpolation due to its low-pass filtering effect \citep{Fraternale2019}. As shown in the figure, in the frequency range $10^{-6}-10^{-4}\,$Hz, $\theta_{k, B_0}$ is predominantly around $90^\circ$ in most of the time period, which may indicate that waves propagate mainly in a direction quasi-perpendicular to the local mean magnetic field. 

During this period, near the maximum of solar cycle 24, the radial distance increases from 103.1 to 106.8 AU, with an average distance of about 105 AU. The azimuthal angle $\phi$ has no primary peak and fluctuates between Parker spiral magnetic field directions $270^\circ$ and $90^\circ$. The elevation angle $\theta$ is also widely distributed and does not lie along 0 degree. These features have been identified as a sector zone \citep{Burlaga2017}. The averaged magnetic field magnitude is about 0.1 nT. The angle between the mean magnetic field $\boldsymbol{B}_0$ and the mean solar wind speed $\boldsymbol{U}_0$ is about 43 degrees during this time period. The averaged solar wind flow velocity $U_p$ is $\sim145$ km/s, averaged density $n_p$ is about 0.002 $cm^{-3}$, and averaged temperature $T_p$ is about 51703 K. We assume that the PUI number density $n_{pui}$ in the inner heliosheath is about $1/4$ of the solar wind proton density, and the temperature $T_{pui}$ is about 180 times the solar wind proton temperature \citep{Zank2010, Zank2018PUI}. The electron number density $n_e$ is assumed to be $n_p + n_{pui}$, and the temperature $T_e$ the same as the solar wind proton temperature $T_p$. Table 1 lists all other relevant parameters used in this paper. As requested, we also include the neutral hydrogen number density $n_H$. Note that $n_H$ is not directly used in our model and is not directly observed by Voyagers. Neutral atom imaging and pickup ion observations, combined with modeling, provide some constraints on the properties of neutral populations. For example, \cite{zhao2019} used a interstellar neutral density of $0.1\,\mathrm{cm}^{-3}$ (which is also the main neutral population in the heliosheath) to fit the pickup ion measurements by New Horizons. However, the exact number of $n_H$ depends critically on other model parameters such as the ionization rate and ionization cavity size. \cite{Bzowski2009} found that $n_H$ at the termination shock is about $0.09\pm0.022\,\mathrm{cm}^{-3}$ based on Ulysses pickup ion observations.    

\begin{table}[htbp]\label{table:up}
\caption{Magnetic field and plasma parameters in the inner heliosheath}
\centering
\begin{tabular}{ccccccccccccccc}
\hline\hline
$|B|$ & $U_{p}$  & $n_{p}$ & $n_{H}$  & $n_{pui}$ & $n_{e}$  & $T_{p}$ & $T_{pui}$ & $T_e$ & $V_A$ & $C_s$ & $V_f$ & $\beta$ & $\omega_{gp}$ & $\omega_{pe}$\\ 
nT & km/s & $cm^{-3}$ & $cm^{-3}$ & $cm^{-3}$ & $cm^{-3}$ & MK & MK & MK & km/s & km/s & km/s & & rad/s & rad/s\\
\hline
	0.1  & 145   & 0.002  &0.1    & 0.0005    & 0.0025   & 0.052    & 9.36  & 0.052  &44  & 164 &170 & 17 & 0.01 & 2821\\
\hline
\end{tabular}
\tablecomments{The Alfv\'en speed $V_A$ and the sound speed $C_s$ are calculated over all species of charged particles (solar wind protons, electrons and PUIs). The estimated fast magnetosonic speed $V_f$ assumes perpendicular propagation only. The plasma beta $\beta$ takes into account the PUI pressure. $\omega_{gp}$ denotes the proton gyrofrequency and $\omega_{pe}$ the electron plasma frequency. }
\end{table}

\begin{figure}
\centering
\includegraphics[width=1.0\linewidth]{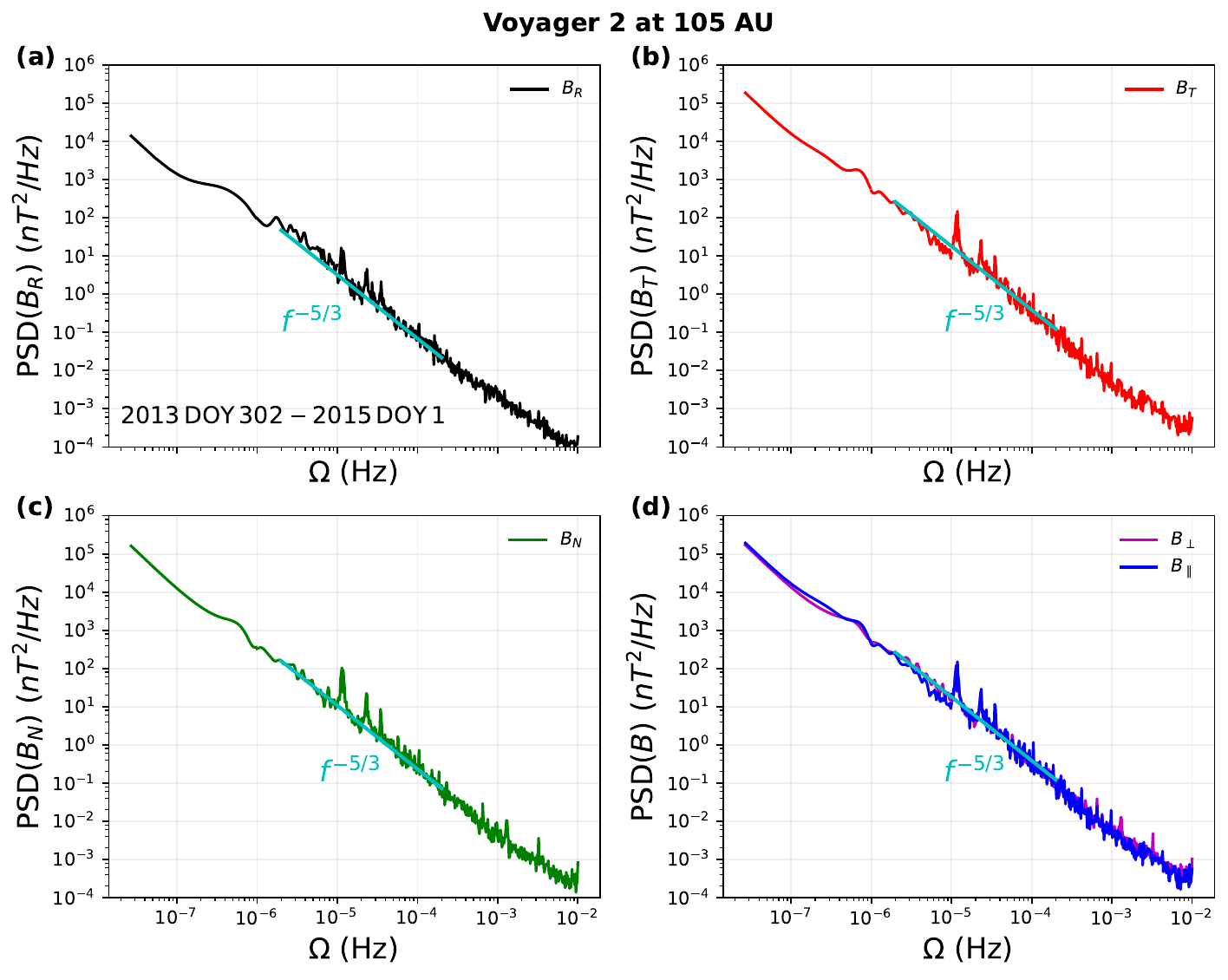}
	\caption{Power spectral density (PSD) of the magnetic field fluctuations observed by Voyager 2 in the inner heliosheath for the same time interval as in Figure \ref{fig:time}. Panels (a)-(d) show the PSDs of the magnetic field components $B_R$, $B_T$, $B_N$, parallel and perpendicular fluctuations, respectively. The parallel spectrum $P_{\parallel}$ represents the PSD along the mean magnetic field $B_\parallel$ direction, and the perpendicular spectrum $P_{\perp}$ is the sum of the PSDs in the two perpendicular directions $B_{{\perp 1}}$ and $B_{{\perp 2}}$. A Kolmogorov spectrum $f^{-5/3}$ is shown as a reference.  }\label{fig:data}
\end{figure}

During this 428-day period, only about 28\% of the 48 s cadence magnetic field data are valid. 
For such unequally spaced time series, we use the method of Lomb-Scargle periodogram \citep{lomb1976, scargle1982, Zech2009, vanderplas2018} to obtain the power spectrum density (PSD) over the frequency space after dividing 
the 428-day time series into fifteen subintervals with an overlap of 50\% between adjacent intervals. 
The final PSD of the magnetic field fluctuations is obtained by averaging the PSDs in these subintervals.
As noted in \citet{Fraternale2019}, the spectrum in the low frequency range (e.g., for frequencies less than $\sim2\times10^{-5}$ Hz) can be well restored by linearly interpolating the data gaps and then performing the Fourier transform on its autocorrelation function.
Figure \ref{fig:data} shows the combined power spectral density (PSD) of the magnetic fluctuations in the inner heliosheath. We perform the Lomb-Scargle periodogram on 48 s cadence magnetic field measurements to obtain the PSD in the frequency range between $10^{-5}$ and $10^{-2}$ Hz. For frequencies below $10^{-5}$ Hz, the analysis is based on the Fourier-transformed autocorrelation function calculated using linear interpolation of 48-second magnetic field data. The spectra of the R, T, N components of the magnetic field are plotted separately.
The parallel and perpendicular spectra are calculated with respect to the mean magnetic field $\boldsymbol{B}_0$ direction. 
We note that there are some spike-like structures in the PSD at around $10^{-5}\,$Hz. We caution that these spikes are unphysical. It can be seen from Figure \ref{fig:time}(g) and (h) that there are no obvious features of specific wave modes or non-propagating structures at around $10^{-5}\,$Hz. The reason these spikes arise in the frequency spectra in Figure \ref{fig:data} is due to the large data gaps and depends also on the spectral estimation technique used to process these data gaps. For the 48-second magnetic field measurements made by Voyager 2, the typical frequency of the large data gaps $\Omega_{gap}$ is about $10^{-5}\,$Hz or a period of $\sim$1 day \citep{Fraternale2019}. For the Lomb-Scargle periodogram used in Figure \ref{fig:data}, large spikes can appear at $\Omega_{gap}$ in the PSD and unphysical power leakage may arise below $\Omega_{gap}$. For frequencies below $\Omega_{gap}$, linear interpolation can overcome this defect and recover low frequencies well as shown in Figure \ref{fig:time}. 
Nevertheless, the observed frequency spectrum has a power-law shape at frequencies above $10^{-6}$ Hz and the power-law index is roughly consistent with $-5/3$. The PSD magnitudes of the T and N components are comparable and much higher than that of the R component. The mean magnetic field $\boldsymbol{B}_0$ is mainly along the T direction, i.e., at an angle of 30$^\circ$ with the T direction, 78$^\circ$ with the N direction, and 63$^\circ$ with the R direction. Therefore, the parallel spectrum $P_{\parallel}$ predominantly comes from the $B_T$ fluctuations, while the perpendicular spectrum $P_\perp$ is mainly contributed by $B_R$ and $B_N$ fluctuations. The bottom right panel shows that the magnetic compressibility, $P_{\parallel}/P_{\perp}$, is much stronger than in the inner heliosphere, where the compressible power is typically $\sim\,10\%$ of the total power, or equivalently $P_{\parallel}/P_{\perp} \sim 1/9$ or $P_{\parallel}/P_{Tr} \sim 0.1$ with $P_{Tr}$ being the total trace spectrum \citep[e.g.,][]{Belcher1971large, smith2006}.
The comparable parallel $P_\parallel$ and perpendicular $P_\perp$ spectra suggests an appreciable level of compressibility, which was also reported in previous studies \citep{Burlaga2017}. 

\subsection{Model of power spectrum with MHD waves}

In this section, we present the spectral modeling results based on the assumption of an isotropic spectrum with a broken power-law shape as discussed in Section \ref{sec:method}. We assume that the PUI number density is about $1/4$ of the solar wind proton number density and the temperature is about 180 times the solar wind proton temperature \citep{Zank2018PUI}. We project the Voyager 2 measured mean bulk speed $\boldsymbol{U}_0$ into the mean field coordinate system (i.e., $\boldsymbol{B}_0$ is along $z$-direction and $\boldsymbol{U}_0$ is on the $x$-$z$ plane), thus $U_x = 97\,$km/s, $U_y = 0$, and $U_z = 106\,$km/s. In addition, the Alfv\'en speed $V_A = 44\,$km/s and the sound speed $C_s = 164 \,$km/s are obtained based on the estimated pickup ions density and temperature \citep{zhao2019}.
We consider the wavenumber spectrum Equation \eqref{eq:3Dspatial} with the parameters $k_0 = 3\times10^{-9}\,$km$^{-1}$, $\alpha = 5/3$, and $P_0 = 2\times10^{-10}\,$nT$^2\,$km$^3$ for both Alfv\'en and fast mode turbulence. These parameters are chosen so that the magnitude and shape of the resulting predicted frequency spectrum roughly agree with the observed magnetic fluctuation PSD shown in Figure \ref{fig:data}. 

\begin{figure}
\centering
\includegraphics[width=0.5\linewidth]{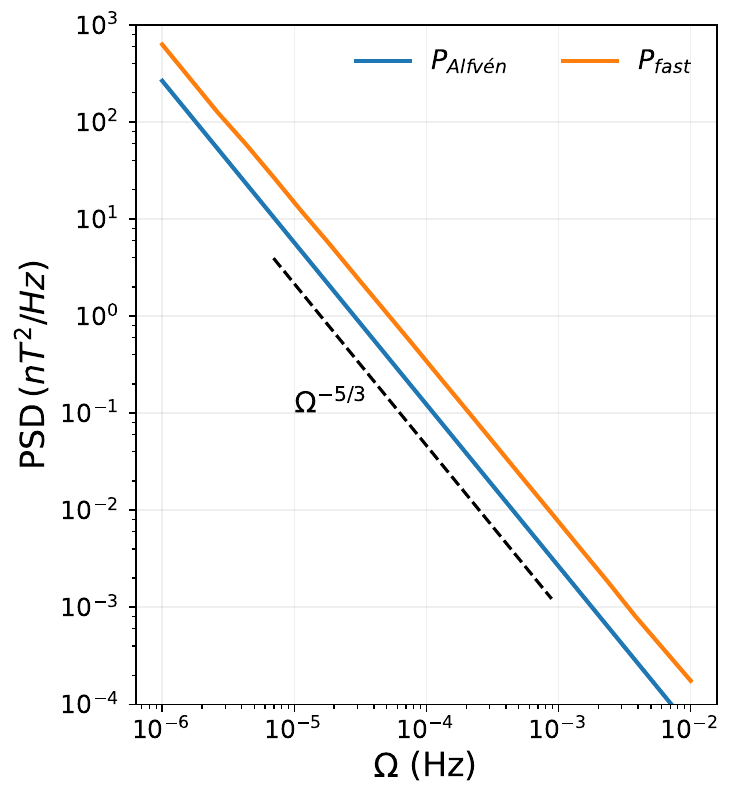}
\caption{The observable power spectra of isotropic Alfv\'en and fast turbulence assuming that both have exactly the same wavenumber spectra. The Voyager 2 solar wind proton measurements and the estimated pickup ion density and temperature in the inner heliosheath are used in the calculation of the dispersion relations for the two wave modes.}\label{fig:fast2}
\end{figure}

Based on Equations \eqref{eq:4F} and \eqref{eq:6F}, we show the observable frequency spectra $P_{obs}(\Omega)$ of the isotropic Alfv\'en and fast mode turbulence in Figure \ref{fig:fast2}. We choose the frequency range of $10^{-6}$--$10^{-2}\,$Hz according to typical Voyager observations \citep[e.g.,][]{Fraternale2019}.
As can be seen from the figure, the spectral indices do not deviate from the $\alpha$ values set in the  wavenumber spectra. However, although the Alfv\'en mode turbulence and the fast mode turbulence have the same wavenumber spectrum, i.e., all parameters in the 3D spatial spectral model are set the same for both, the observable spectra of the Alfv\'en mode and the fast mode are different in terms of the spectral power. This illustrates the break down of the standard Taylor hypothesis as the spacecraft observed frequency spectrum cannot be directly transformed from the wavenumber spectrum.   
From Figure \ref{fig:fast2}, MHD fast-mode turbulence has higher observed power due to its higher propagation speed. As discussed before, the physical meaning of Equation \eqref{eq:4Dspectra} is that the turbulence observed at a given frequency is a superposition of fluctuations with various wavevectors and flow-frame frequencies, all Doppler shifted to the same observed frequency.
In the spacecraft frame, fast modes propagate at a speed of $U_x + \sqrt{V_A^2 + C_s^2} \sim$ 267$\,$km/s in $x$-direction and $U_z + \max(V_A, C_s) \sim$ 270$\,$km/s in $z$-direction, while Alfv\'en waves propagate at $U_x = $ 97$\,$km/s in the $x$-direction and $U_z + V_A \sim$ 150$\,$km/s in $z$-direction.
Compared to Alfv\'en waves, fast mode waves with longer wavelengths can be Doppler-shifted to the same observed frequency due to their larger propagation speed.
As the longer-wavelength modes contain stronger fluctuations, fast-mode turbulence is Doppler boosted to a higher power than Alfv\'enic turbulence.
This is analogous to the results shown in \citet{Zhao2024} and \citet{Goldstein1986}, where the difference in the propagation speed of outward and inward Alfv\'en waves can lead to an apparent imbalance in the observed fluctuations close to the Sun.
\begin{figure}
\centering
\includegraphics[width=0.5\linewidth]{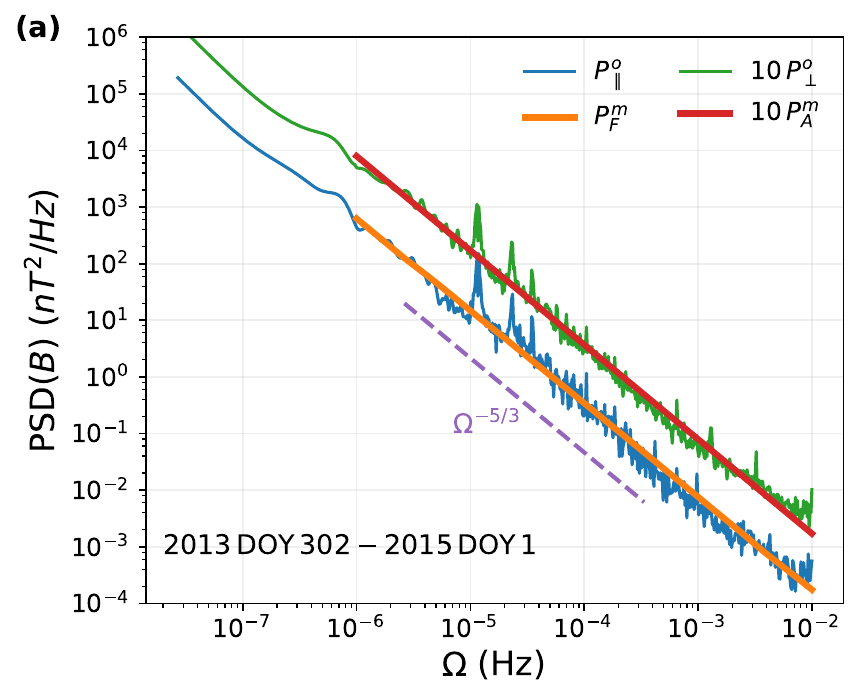}%
\includegraphics[width=0.5\linewidth]{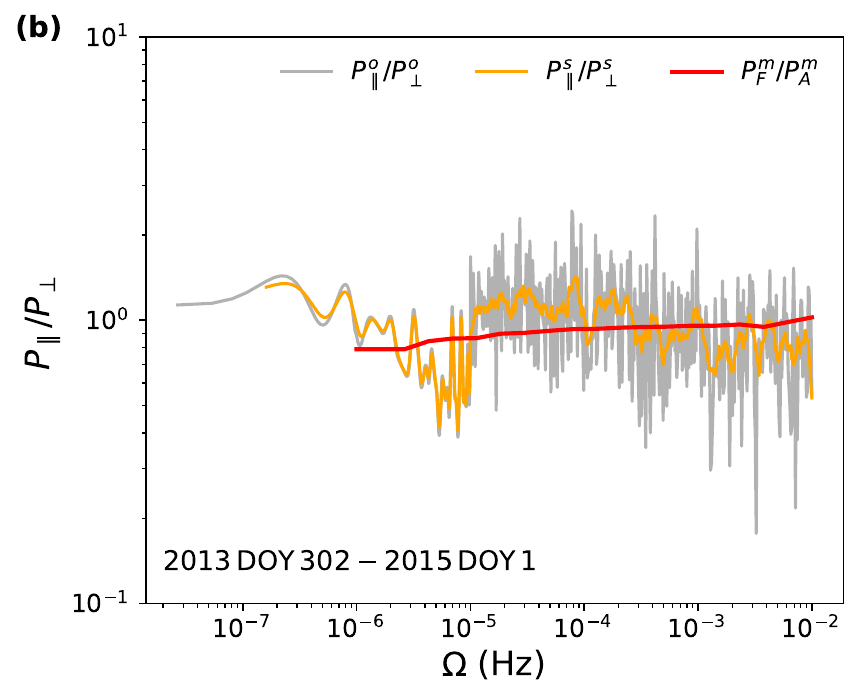}
	\caption{Panel (a) shows the comparison between the modeled observable spectrum of the isotropic Alfv\'en turbulence $P^m_A$ and Voyager 2 measured perpendicular spectrum $P^o_\perp$, and between the observable spectrum of the isotropic fast mode $P^m_F$ and the measured parallel spectrum $P^o_\parallel$. The measured parallel and perpendicular spectra ($P^o_\parallel$ and $P^o_\perp$) are the same as in Figure \ref{fig:data}. For ease of presentation, $P^o_\perp$ and $P^m_A$ have been multiplied by 10. Panel (b) shows a comparison of the magnetic compressibility, i.e., the ratio between the parallel and perpendicular spectra $P_\parallel/P_\perp$. The grey curve shows the Voyager 2 measured value, the orange curve denotes the smoothed compressibility, and the red curve shows the ratio between the modeled observable spectra for fast turbulence and Alfv\'en mode turbulence.}\label{fig:model}
\end{figure}
If turbulence in the heliosheath is indeed a superposition of isotropic Alfv\'en and fast modes, our method enables the calculation of the power fraction in these two components, as indicated by the spectral power normalization parameter $P_0$.
Specifically, we use the MHD fast mode to approximately characterize the observed compressible fluctuations and the Alfv\'en mode to represent the observed incompressible fluctuations.  
Figure \ref{fig:model} (a) shows the modeled spectra for isotropic fast turbulence $P_F^m$ and isotropic Alfv\'en turbulence $P_A^m$, compared with the observed parallel $P^o_\parallel$ and perpendicular $P^o_\perp$ spectra, respectively.
$P^o_\perp$ and $P_A^m$ have been multiplied by a factor of 10 for presentation purposes. We note that the spectral indices of the calculated observable spectra for the isotropic Alfv\'en turbulence and fast mode turbulence still retain the Kolmogorov shape consistent with their wavenumber spectra ($\alpha = 5/3$).
From Figure \ref{fig:data}, we know that the ratio of the observed parallel to perpendicular power spectra is about 1. However, if we assume that the Alfv\'en and fast modes have the same power normalization parameter $P_0$, then the power ratio between their observable spectra $P_{Fast}/P_{Alfven}\sim3$ as shown in Figure \ref{fig:fast2}. Therefore, in order to make their observable spectra consistent with the actual observed $P_\parallel$ and $P_\perp$, we find that the value of $P_0$ for the Alfv\'en mode needs to be three times that of the fast mode component. Specifically, we use $P_0^A = 6\times10^{-10}\,$nT$^2\,$km$^3$ and $P_0^F = 2\times10^{-10}\,$nT$^2\,$km$^3$ to quantitatively model the observed perpendicular $P^o_\perp$ and parallel $P^o_\parallel$ fluctuation spectra. In panel (b), we show the modeled compressibility compared to the observed magnetic compressibility, which is expressed as the power ratio between parallel and perpendicular fluctuations. The grey curve indicates the observed compressibility calculated by $P^o_\parallel/P^o_\perp$, the orange curve shows the smoothed ratio as a result of 10 data points moving average, and the red curve shows the power ratio between the modeled fast turbulence and the modeled Alfv\'en turbulence. 
Since $P_0^A = 3P_0^F$ is required to obtain the consistent compressibility as measured by Voyager 2, it means that the actual ratio between compressible fluctuation power and incompressible fluctuation power is $\sim 1/3$, which is noticeably higher than the nominal value of $1/9$ for the solar wind near the Earth \citep{Belcher1971large, pine2020solar}, but not as high as the ratio of 1 suggested by direct measurements \citep{Burlaga2006, Burlaga2009}.

\subsection{Model of power spectrum with dispersive waves}

\begin{figure}
\centering
\includegraphics[width=0.5\linewidth]{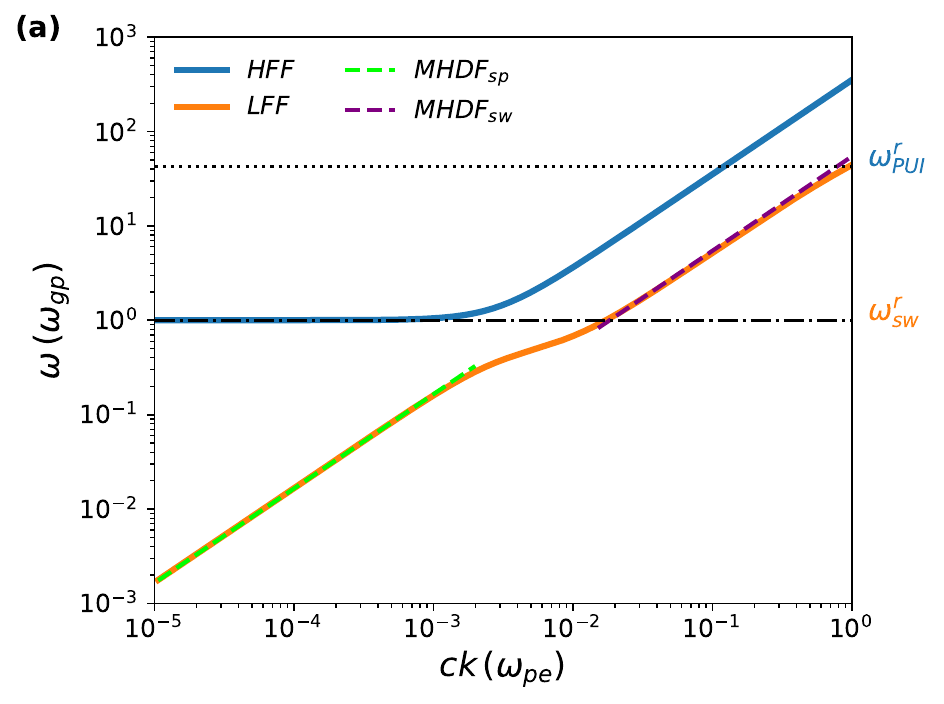}%
\includegraphics[width=0.46\linewidth]{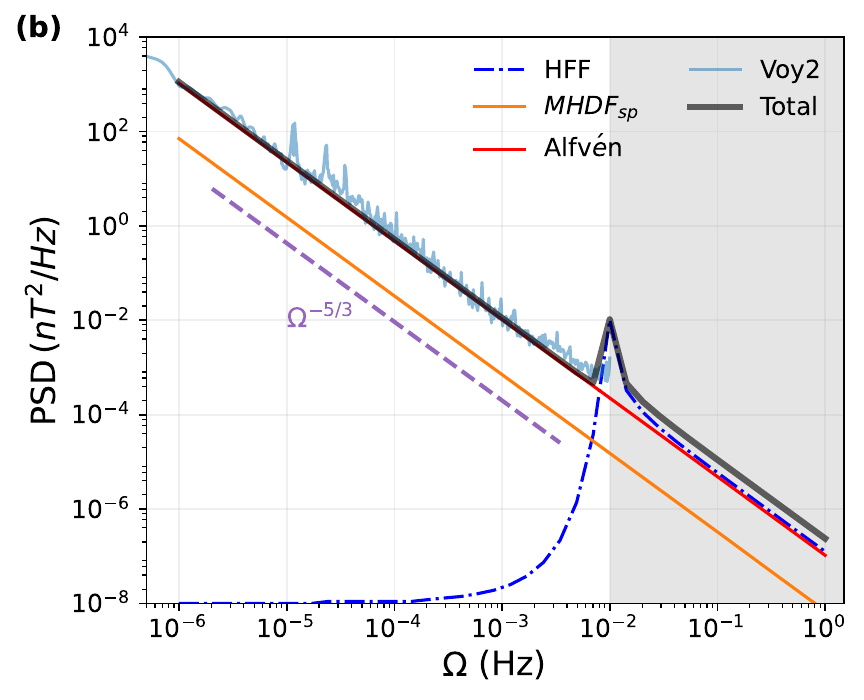}
	\caption{Panel (a) shows the dispersion relations of the PUI high-frequency fast mode (HFF, blue solid line) and solar wind ion low-frequency fast mode (LFF, orange solid line) from a three-fluid model \citep{Zieger2015}. The purple dashed line MHDF$_{sw}$ shows the MHD fast mode considering only solar wind ions. The green dashed line MHDF$_{sp}$ denotes the MHD fast mode considering both PUIs and solar wind ions. The plasma frame frequency $\omega$ is normalized to the proton gyrofrequency $\omega_{gp}$, and the wave number $k$ is normalized to the electron inertial length $c/\omega_{pe}$. The dashed-dotted horizontal line represents the cutoff frequency of the HFF mode and the resonance frequency of the LFF mode. The dotted horizontal line represents the resonance frequency of the HFF mode. Panel (b) shows the observable frequency spectrum for each mode. The solid black line shows the total frequency spectrum calculated from the sum of the HFF, MHDF$_{sp}$, and isotropic Alfv\'en modes, which is consistent with the PSD observed by Voyager 2 in the frequency range $10^{-6}$--$10^{-2}$ Hz (cyan curve). The grey shaded area indicates the region that cannot be seen by Voyager 2 due to the low signal-to-noise ratio in this frequency range.}\label{fig:f1}
\end{figure}

As the hotter PUIs and relatively ``cold'' solar wind ions may introduce separate fast magnetosonic modes based on the multifluid description \citep[e.g.,][]{Zank2014, Zieger2015}, we also consider the observable frequency spectrum for these two types of dispersive waves. Figure \ref{fig:f1} (a) shows the dispersion relation of the two dispersive fast branches HFF and LFF from Equation \eqref{eq:dispersion} together with two MHD fast modes (MHDF) for comparison. The phase velocities of the two MHDF modes are calculated by considering only the solar wind ions (MHDF$_{sw}$) and by considering both PUIs and solar wind ions (MHDF$_{sp}$), respectively. In plotting Figure \ref{fig:f1}, the solar wind ion parameters are based on observations from Voyager 2 in the helisheath (Figure \ref{fig:time}). The pickup ion and electron parameters are based on theoretical assumption and numerical simulation results \citep{Zank2018PUI, Zieger2020} and all are listed in Table 1. As discussed in Section \ref{sec:method}, we consider perpendicular propagation only for both HFF and LFF modes. The higher-frequency fast branch (HFF, blue solid line) is due to the higher-temperature pickup ions and the lower-frequency fast mode branch (LFF, orange solid line) is due to the lower-temperature core solar wind ions. The LFF branch has a resonance frequency at the proton gyrofrequency $\omega_{gp}$.
An interesting feature is that the HFF branch has a cutoff frequency also at $\omega_{gp}$, i.e., this branch cannot exist below the cutoff frequency $\omega_{gp}$. The HFF branch is actually the ion Bernstein wave mode, i.e., an electrostatic ion cyclotron wave propagating perpendicular to the magnetic field. In contrast, the solar wind ion-driven LFF branch is very similar to the MHD fast mode by considering different ion species. For instance, the LFF mode at the small wavenumbers ($k\le$ 0.003 $\omega_{pe}/c \sim 3\times10^{-5}\,$km$^{-1}$) basically follows the dispersion relation of MHDF$_{sp}$, where the phase speed is calculated by $\sqrt{V_A^2+C_s^2}$ for the perpendicular propagation. The phase speed for MHDF$_{sp}$ mode, $V^{sp}_{f} \simeq 168\,$km/s, includes both PUIs and solar wind ions contributions. At large wavenumbers ($k\ge$ 0.01 $\omega_{pe}/c \sim 10^{-4}\,$km$^{-1}$), the LFF mode follows the MHD fast mode MHDF$_{sw}$, where the phase speed, $V^{sw}_{f} \simeq 56\,$km/s, is calculated from solar wind ions only.        
In Figure \ref{fig:f1} (b), we compute the spacecraft observable frequency spectra for each mode based on Equation \eqref{eq:18F}. 
The presence of two dispersive fast mode branches causes extra complications to spectral modeling. Physically, it is not clear how fluctuation power is divided between the two branches of fast modes, so assumptions have to be made in the present work.
Here, we make the simplest assumption that both branches have the same power spectrum in wavenumber space, i.e., $P^{lff}_0 = P^{hff}_0 = 2\times10^{-10}$ and $\alpha = 5/3$ in Equation \eqref{eq:multi3D}. The LFF mode at small wavenumbers can be replaced by the MHDF$_{sp}$ and the spectrum observed at large frequencies should be dominated by the HFF mode due to its larger phase velocity. Here, we derive the spacecraft frame frequency spectra of the HFF (blue dashed-dotted line) and the MHDF$_{sp}$ (orange line) for simplicity. The difference between the MHDF$_{sp}$ and the isotropic MHD fast mode shown in Figure \ref{fig:fast2} is that the MHDF$_{sp}$ only has perpendicular propagation ($\boldsymbol{k}$ has only $k_x$ and $k_y$) to be consistent with the LFF mode, while the isotropic MHD fast mode in Figures \ref{fig:fast2} and \ref{fig:model} has 3D 
wavevector ($k_x$, $k_y$, and $k_z$). 
It can be seen that the MHDF$_{sp}$ still retains the $-5/3$ spectral shape as its wavenumber spectrum.
We also show the frequency spectrum of the isotropic Alfv\'en mode taken from Figure \ref{fig:model}. The total PSD (solid black line) is plotted as the sum of the three (HFF, MHDF$_{sp}$, and Alfv\'en), which exhibits a clear spectral bump near the proton gyrofrequency that is caused by the HFF mode. At low frequency range ($10^{-6}-10^{-2}\,$Hz), the theoretical predicted total PSD is consistent with the PSD measured by Voyager 2 (cyan line). In fact, the HFF mode contributes to the observed total PSD only at frequencies above the proton gyrofrequency, and its power can be negligible below it, which is consistent with the cutoff behavior shown in Panel (a). The spectral bump at the proton gyrofrequency is also present in the simulation by \cite{Zieger2020}, which is due to the ion-ion resonance instability that drives the inverse cascade of turbulence.  
Since the highest cadence of magnetic field measurements made by Voyager 2 in the heliosheath is 48 seconds, the spectrum can be measured only below the Nyquist frequency of about 0.01 Hz or less than $10^{-4}\,$km$^{-1}$. The gray shaded area in Panel (b) represents the unexplored region that cannot be seen by Voyager 2 because of the low signal-to-noise in this frequency range. We also note that the presence of the spectral bump shown in the total observable PSD depends on how we distribute the fractional power among these three modes. Panel (b) shows the resulting observable spectrum with a fractional power ratio of Alfv\'en, HFF, and MHDF$_{sp}$ as 3:1:1 is consistent with the observed PSD at the frequency range $10^{-6}-10^{-2}\,$Hz.  

We note that the nature of the LFF mode is that it is dominated by the solar wind ions at larger wavenumbers, while it reduces to a single-fluid-like fast mode at smaller wavenumbers. This is more clearly described in Figure 5(a). At frequencies larger than the resonance frequency of the LFF mode or the cutoff frequency of the HFF mode (i.e., $\omega^r_{sw}$), the dispersion of the LFF mode is roughly consistent with the MHD fast mode that considers solar wind ions only (e.g., MHDF$_{sw}$). But at lower frequencies ($\omega<\omega^r_{sw}$), the LFF mode reduces to a single-fluid-like fast mode including the contribution of both solar wind ions and pickup ions (e.g., MHDF$_{sp}$). Therefore, the PSD at the current spacecraft resolution does not necessarily reflect solar wind ion dominated turbulence only, but rather the effects of a single-fluid-like system with contributions from both solar wind ions and pickup ions. Most likely, the solar wind ion dominated turbulence is important for PSD near the highest observable frequency ($\sim10^{-2}\,$Hz), and turbulence is single-fluid-like at lower frequencies. PUI-dominated turbulence is likely important for even higher frequencies. It is possible that that there are significant unobserved power at higher frequencies predicted by \cite{Zieger2015, Zieger2020}, but we are not aware of available data at higher frequencies by Voyagers to verify this.

Simulations by \citet{Zieger2020} suggested that the breakdown of Taylor's hypothesis can also cause the spectral index of the observed frequency spectrum to deviate from the spectral index of the corresponding wavenumber spectrum. They show that a wavenumber spectrum of $\sim k_{\perp}^{-4}$ can produce the observed spectrum of $\sim f^{-5/3}$. This is possible for dispersive waves because the spectral power at a higher wavenumber is more strongly enhanced by the Doppler shift, making the frequency spectrum flatter than the wavenumber spectrum. However, this effect is not included in our results because the wavenumber regime of strong dispersion corresponds to high frequencies that are barely resolved by Voyagers data. 
Another complication is that the decomposition between the high-frequency and low-frequency fast modes is not well understood. Since this decomposition depends on the wavenumber, the possibilities are almost endless and likely cannot be constrained by observations. A more detailed theoretical description of the full 4D spectrum is needed for further progress in this direction.

\section{Discussions and conclusions}
In this paper, we discuss the compressive properties of turbulence in the inner heliosheath. A possible origin of the compressible turbulence observed in the heliosheath are the solar wind fluctuations upstream of the HTS. These fluctuations interact with the quasi-perpendicular HTS and are transmitted downstream with an enhanced PSD \citep{Zank2021ApJ}.
Fast mode turbulence in the heliosheath can be further transmitted across the heliopause to generate compressible turbulence in the interstellar medium \citep{Zank2017VISM}, as seen by the Voyagers spacecraft \citep{Zhao2020, Burlaga2022}.
The standard Taylor's hypothesis is commonly used to interpret in-situ observed turbulent signals throughout the heliosphere and in the interstellar medium. However, the implicit application of Taylor's hypothesis can lead to incorrect interpretation of fluctuation measurements, especially when the characteristic propagation speed of the fluctuations is larger or comparable to the flow speed, a condition that applies to both the inner heliosphere and inner heliosheath regions. We introduce a 4D frequency-wavenumber spectral modeling method to overcome this caveat. A 4D $\omega$-$\boldsymbol{k}$ spectrum can be used to extend the standard Taylor's hypothesis through the inclusion of the fluctuation frequencies.
We find that the inclusion of temporal (or frequency space) changes of the fluctuations has important implications for the interpretation of turbulence measurements in the inner heliosheath by Voyager 1 and 2 or future missions, such as New Horizons that is expected to cross the HTS in the next few years and Interstellar Probe. Our results demonstrate that 

1) In-situ observations in the heliosheath favor the identification of fast modes over Alfv\'en modes, which leads to a much higher magnetic compressibility observed in the heliosheath. We caution that this is an observational bias and the true compressibility needs to be revisited through spectral modeling.

2) Assuming that the wavenumber spectrum is Kolmogorov-like, equipartition in the isotropic fast and Alfv\'enic fluctuations will lead to an observed power ratio between the two of $\sim$3:1.
Since Voyager observations tend to find comparable power in compressible and incompressible magnetic fluctuations in the heliosheath, our results suggest that turbulence is less compressible than previously thought. In other words, the fractional ratio of the compressible fluctuations may account 25\% of the total fluctuations rather than 50\% as suggested by Voyager observations. 

3) Hot pickup ions and relatively cold solar wind ions may introduce two fast wave mode branches, namely a high-frequency fast mode (HFF) due to the pickup ion component and a low frequency fast mode (LFF) driven by solar wind ions. Both modes are dispersive waves and can affect the observed spectral shape of turbulence, leading to discrepancies between the wavenumber spectrum and the observed 1D frequency spectrum. The pickup ion-driven HFF mode has a cut-off frequency at the proton gyrofrequency and may lead to an observable spectral bump near the proton gyrofrequency depending on the power partitioning among different wave modes. 

We caution that the multi-fluid fast mode waves, HFF and LFF, are derived based on an isotropic distribution of PUIs and only admit perpendicular propagation \citep{Zieger2015}. Since PUIs in the heliosheath undergo pitch-angle scattering, the isotropic assumption may not be valid. A more general fluid model is developed by \cite{Zank2014} with collisionless heat flux and viscosity included, and waves modes can experience damping because of it. However, in the perpendicular propagation limit, \citet{Zank2014} and \citet{Zieger2015} give the same results, i.e., PUI fast magnetosonic wave and solar wind ion fast wave. Using a more general form of the dispersive fast wave turbulence \citep{Zank2014} in any propagation direction will be considered in future work. In addition, the isotropic turbulence spectral model (i.e., spectral power depends on the magnitude of wavenumber only) we used in this work may not be accurate, though it is justifiable due to the weak magnetic field in the heliosheath. Anisotropic turbulence models may be worth considering for comparison.

We also assume that fast mode turbulence only contributes to the compressive component of the spectral power. This is certainly an oversimplification as the fast mode is expected to contain both compressive and incompressible magnetic field fluctuations. If the fast mode polarization conditions are taken into account, we would expect the power ratio between Alfv\'en and fast-mode turbulence to decrease. However, the qualitative conclusion still holds that the compressibility is overestimated when using standard Taylor's hypothesis to interpret the data.
Furthermore, the present model does not include compressible mirror modes that may exist downstream of the HTS \citep{Liu2007}. Due to the high plasma beta in the heliosheath, it is possible that mirror modes do contribute to the observed compressive fluctuations. However, including mirror modes will introduce further complications to the modeling of the observed spectra, and there are no direct observations of temperature anisotropy in the heliosheath. We therefore defer it to a future study.

Another caveat in this work is the assumed PUI number density and temperature in the heliosheath. Since PUIs are not directly measured by Voyagers, there are uncertainties associated with parameters related to PUIs. The most important parameter is the PUI pressure as it dominates the sound speed calculation. The results are less sensitive to PUI number density since it is expected to be smaller than the relatively ``cold'' solar wind ions.

To summarize, the 4D power spectrum modeling is a decomposition of different wave modes following their respective dispersion relations. To go beyond Taylor's hypothesis, the key is to use the 4D $\omega$-$\boldsymbol{k}$ spectrum to model the observed 1D reduced spectrum. 
We emphasize that a quantitative spectral modeling of the measured fluctuations is critical for the interpretation of in-situ turbulence data without Taylor's hypothesis. Future turbulence modeling efforts in the heliosheath will be of great importance \citep{opher2023solar}.

\section*{\leftline{Acknowledgement}}
\begin{acknowledgments}
We acknowledge the partial support of the NSF EPSCoR RII-Track-1 Cooperative Agreement OIA-1655280, NASA awards 80NSSC20K1783, 80NSSC23K0415, 80NSSC23K0660, and a NASA IMAP subaward under NASA contract 80GSFC19C0027, and a NASA Heliospheric DRIVE Center award SHIELD 80NSSC22M0164. 	
\end{acknowledgments}

\appendix
\section{Calculation of the frequency spectrum}

Here, we describe how the integration is computed to obtain the frequency spectrum.
For Alfv\'enic turbulence, the 2D integration is relatively straightforward. The integration limit is $-\infty$ to $\infty$ for both $k_y$ and $k_z$.
The integrand is symmetric about $k_y = 0$, so the integral can be written as
\begin{equation}
  P_{obs}(\Omega) = 2\int_{-\infty}^{\infty} \int_{0}^{\infty} \frac{1}{U_x} P\left( \sqrt{\left( \frac{k_z U_z + |k_z V_A| - \Omega}{U_x} \right)^2 + k_y^2 + k_z^2} \right) dk_y dk_z
\end{equation}
Numerically, since most power in the 3D spectrum is due to small-wavenumber fluctuations, the integral may be approximated with finite limits.

For fast-mode turbulence, we convert the 3D integral to 2D by integrating over $k_y$ using the delta function, to obtain
\begin{equation}\label{eq:xz}
  P_{obs}(\Omega) = \int \frac{1}{|\partial(k V_f)/\partial k_y|_{k_{y0}}} P\left( \sqrt{k_x^2 + k_{y0}^2(k_x, k_z, \Omega) + k_z^2} \right) dk_x dk_z,
\end{equation}
where $k_{y0}$ is determined by the argument of the delta function, i.e.,
\begin{equation}\label{eq:ky0}
  k_x U_x + k_z U_z + \sqrt{k_x^2 + k_{y0}^2 + k_z^2} V_f(k_x, k_{y0}, k_z) - \Omega = 0,
\end{equation}
to be solved numerically for $k_{y0}$.
There are two solutions with opposite signs and they contribute equally to the integral, so we can simply keep the positive one (denoted as $k_{y+}=|k_{y0}|$) and double the result.
The integration limits for $k_x$ and $k_z$ are not $-\infty$ to $\infty$ because Equation \eqref{eq:ky0} does not have a real solution when $|k_x|$ or $|k_z|$ is too large.
In fact, the left hand side of Equation \eqref{eq:ky0} has a minimum at $k_y = 0$.
Essentially, the integration domain corresponds to the region where the minimum is not positive.

\begin{figure}
\centering
\includegraphics[width=0.5\linewidth]{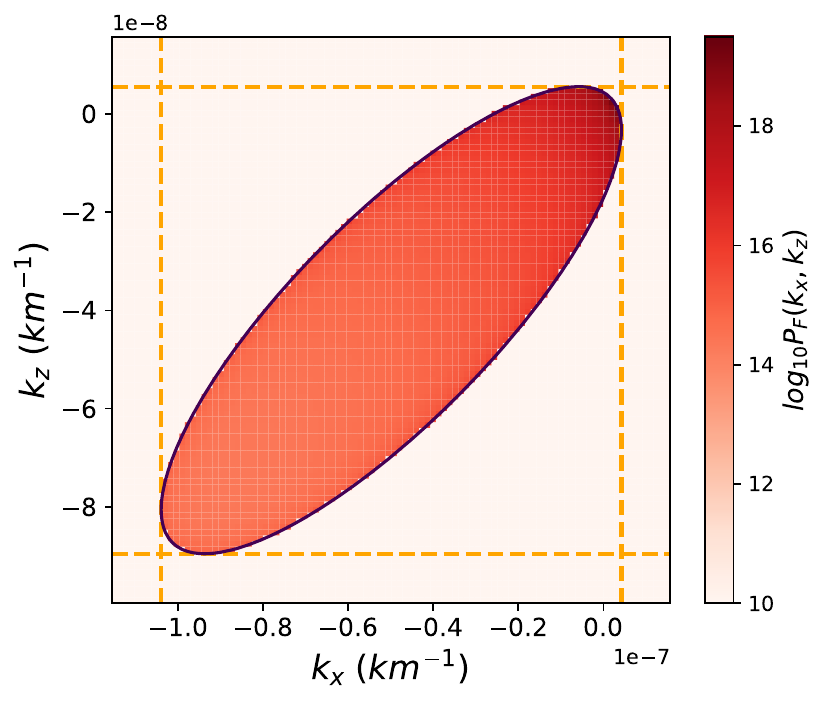}
\caption{The integrand for calculating the fast-mode turbulence spectrum.}\label{fig:integ}
\end{figure}

The integration domain is illustrated in Figure \ref{fig:integ}, where the log of the integrand for $\Omega=10^{-6}$ is plotted in the $k_x-k_z$ plane and we set the integrand to zero outside the domain.
Based on these, the integral is expressed as follows,
\begin{equation}
  P_{obs}(\Omega) = \int_{k_{z1}}^{k_{z2}} \int_{k_{x1}(k_z)}^{k_{x2}(k_z)} \frac{2}{|\partial(k V_f)/\partial k_y|_{k_{y+}}} P\left( \sqrt{k_x^2 + k_{y+}^2(k_x, k_z, \Omega) + k_z^2} \right) dk_x dk_z.
\end{equation}
We integrate $k_x$ first. Given $k_z$ (and $\Omega$), the limits of $k_x$ are found by solving Equation \eqref{eq:ky0}, and letting $k_{y0} = 0$. It can be shown that there are two real solutions in general when $k_z$ is also in the proper range, corresponding to the lower and upper limits of the integration $k_{x1}$ and $k_{x2}$.
The limits in $k_z$ are determined, again, based on Equation \eqref{eq:ky0} by requiring
\begin{equation}
  \min\left[ k_x U_x + k_z U_z + \sqrt{k_x^2 + k_z^2} V_f(k_x, k_z) - \Omega \right] \equiv F(k_z) = 0.
\end{equation}
The minimum is achieved where the derivative equals zero, i.e.,
\begin{equation}
  \frac{\partial}{\partial k_x}\left[ k_x U_x + k_z U_z + \sqrt{k_x^2 + k_z^2} V_f(k_x, k_z) - \Omega \right] = 0.
\end{equation}
This yields the relation between $k_x$ and $k_z$ for the minimum, and thus $F(k_z) = 0$ can be solved numerically for the limits $k_{z1}$ and $k_{z2}$.

\bibliography{freq-wave}
\bibliographystyle{aasjournal}



\end{document}